\pdfoutput=1

\documentclass[11pt]{article}

\usepackage[]{ACL2023}

\usepackage{times}
\usepackage{latexsym}

\usepackage[T1]{fontenc}

\usepackage[utf8]{inputenc}

\usepackage{microtype}

\usepackage{inconsolata}
\usepackage{url}            
\usepackage{booktabs}       
\usepackage{amsfonts}       
\usepackage{nicefrac}       
\usepackage{microtype}      
\usepackage{xcolor}         
\usepackage{multirow}
\usepackage{amsmath}
\usepackage{graphicx}
\usepackage{colortbl}
\usepackage{arydshln}
\usepackage{wrapfig}
\usepackage{subcaption}
\usepackage{mdframed}
\usepackage{pifont}
\usepackage{float}
\usepackage{placeins}

\definecolor{my_green}{RGB}{0, 153, 51}
\definecolor{my_red}{RGB}{204, 0, 0}
\renewcommand{\checkmark}{\textcolor{my_green}{\ding{51}}} 
\newcommand{\crossmark}{\textcolor{my_red}{\ding{55}}} 

\title{SEA-SQL: Semantic-Enhanced Text-to-SQL with Adaptive Refinement}

\author{Chaofan Li$^{1}$, ~Yingxia Shao$^{1}$\footnotemark[1], ~Yawen Li$^{1}$, Zheng Liu$^{2}$\\
  1: Beijing University of Posts and Telecommunications \\
  2: Beijing Academy of Artificial Intelligence \\
  \texttt{\{cfli, shaoyx\}@bupt.edu.cn} \quad
  \texttt{warmly0716@126.com} \quad
   \texttt{zhengliu1026@gmail.com} \\
}

\begin{document}
\maketitle
\begin{abstract}
Recent advancements in large language models (LLMs) have significantly contributed to the progress of the Text-to-SQL task. A common requirement in many of these works is the post-correction of SQL queries. However, the majority of this process entails analyzing error cases to develop prompts with rules that eliminate model bias. And there is a weakness of execution verification for SQL queries. In addition, the prevalent techniques primarily depend on GPT-4 and few-shot prompts, resulting in expensive costs. To investigate the effective methods for SQL refinement in a cost-efficient manner, we introduce Semantic-Enhanced Text-to-SQL with Adaptive Refinement (SEA-SQL), which includes Adaptive Bias Elimination and Dynamic Execution Adjustment, aims to improve performance while minimizing resource expenditure with zero-shot prompts.
Specifically, SEA-SQL employs a semantic-enhanced schema to augment database information and optimize SQL queries.
During the SQL query generation, a fine-tuned adaptive bias eliminator is applied to mitigate inherent biases caused by the LLM.
The dynamic execution adjustment is utilized to guarantee the executability of the bias eliminated SQL query.
We conduct experiments on the Spider and BIRD datasets to demonstrate the effectiveness of this framework. The results demonstrate that SEA-SQL achieves state-of-the-art performance in the GPT3.5 scenario with 9\%-58\% of the generation cost. Furthermore, SEA-SQL is comparable to GPT-4 with only 0.9\%-5.3\% of the generation cost. Our code is available at \href{https://github.com/545999961/SEA-SQL}{\textit{https://github.com/545999961/SEA-SQL}}.
\end{abstract}

\section{Introduction}

Relational databases play a pivotal role in information technology by offering a structured and organized framework for data storage. Through tables, rows, and columns, data is efficiently organized, ensuring clarity and ease of management. However, users typically interact with databases through SQL queries, which is a barrier for those unfamiliar with SQL syntax. To alleviate the gap between natural language and SQL queries, the Text-to-SQL task is proposed \cite{DBLP:conf/ijcai/CaiXZYLL18}. Specifically, given a natural language question, Text-to-SQL aims to convert the question into a SQL query, making it more convenient for users to interact with database.

Traditional sequence-to-sequence methods mainly focus on fine-tuned encoder-decoder models \cite{DBLP:conf/acl/RaiWZY23, DBLP:conf/aaai/Li00023, DBLP:conf/slt/ZengPH22}, which are optimized for the encoding process and decoding process, including schema linking \cite{DBLP:conf/acl/LiuYZGZL21, DBLP:journals/corr/abs-2103-04399}, building intermediate results between the natural language and SQL \cite{DBLP:journals/corr/abs-1810-05237, DBLP:conf/acl/WangSLPR20, DBLP:conf/emnlp/GanCXPWDZ21}, decoding constraints \cite{DBLP:conf/acl/YinN17, DBLP:conf/emnlp/ScholakSB21, DBLP:conf/acl/SuhrCSL20}, etc.
As databases continue to expand in size, they often encompass a large number of tables and columns, complex inter-table relationships, diverse data types, and vast amounts of data. Consequently, user queries tend to exhibit complex structures, such as nested queries, aggregations, joins, or multiple conditions \cite{DBLP:conf/nips/LiHQYLLWQGHZ0LC23}. Both factors increase the complexity of inputs for the Text-to-SQL task. Sequence-to-sequence models have inherent limitations due to the small size of their parameters. As the complexity of input increases, these models may struggle to encode all necessary information efficiently, leading to performance bottlenecks.

Recently, with the rise of LLMs, there has been a significant advancement in natural language understanding, contextual awareness, and scalability. LLMs are capable of effectively managing and interpreting complex queries and schema structures. This has made prompt-based Text-to-SQL approaches more accessible and effective. By simply providing prompts, LLMs can effortlessly get SQL queries and obtain outstanding performance \cite{DBLP:journals/corr/abs-2303-13547, DBLP:journals/corr/abs-2305-11853, DBLP:journals/corr/abs-2307-07306, DBLP:conf/nips/PourrezaR23, DBLP:journals/pvldb/GaoWLSQDZ24, DBLP:journals/dase/ZhouSL24}.
However, previous works in employing LLMs for Text-to-SQL prompting have three major limitations.
(1) \textbf{Inherent model bias}: When generating SQL queries, models may exhibit biases due to patterns present in their training data. Such biases might manifest in specific preferences within the generated queries. For example, GPT-3.5 is case-insensitive and may inadvertently convert an uppercase value from the input into a lowercase value in the SQL query \cite{DBLP:journals/corr/abs-2303-13547}. This behavior reflects the model's inherent biases and can lead to inaccuracies in query formation.
(2) \textbf{Unexecutable SQL}: Although SQL is generated by the LLM, they lack the capability to verify the executability of the final SQL query. When executed, these queries may encounter various errors, such as column mismatches or runtime errors. This limitation underscores the need for additional validation steps to ensure the generated SQL is functional and accurate.
REACT \cite{DBLP:conf/iclr/YaoZYDSN023} demonstrates that the LLM may generate illusory answers in question answering (QA) scenarios without external retrieval capabilities, incorporating retrieval into LLM generation enhances the accuracy and reliability of answer generation. Similarly, the LLM in the Text-to-SQL domain can also generate non-executable SQL queries. Therefore, the absence of external actions such as SQL execution limits the assurance of the executability of the generated SQL.
(3) \textbf{Expensive inference cost}: recent works \cite{DBLP:conf/nips/PourrezaR23, DBLP:journals/pvldb/GaoWLSQDZ24, DBLP:journals/corr/abs-2312-11242} focus on leveraging GPT-4 for SQL generation. However, utilizing GPT-4 comes at a significant expense. Furthermore, most GPT-4 based methods \cite{DBLP:conf/nips/PourrezaR23, DBLP:journals/pvldb/GaoWLSQDZ24, DBLP:journals/corr/abs-2312-11242} rely on few-shot prompts, which leads to longer inputs and higher computational costs. For instance, DAIL-SQL \cite{DBLP:journals/pvldb/GaoWLSQDZ24} requires nearly \$0.1 per SQL query, MAC-SQL \cite{DBLP:journals/corr/abs-2312-11242} requires \$0.2 per SQL query, and DIN-SQL \cite{DBLP:conf/nips/PourrezaR23} requires nearly \$0.8 to generate each SQL query. If these methods are employed to generate thousands or even tens of thousands of SQL statements, it would be a catastrophic cost for ordinary users.

To mitigate these issues, we introduce Semantic-Enhanced Text-to-SQL with Adaptive Refinement (SEA-SQL), an innovative framework based on zero-shot prompts and GPT-3.5 to accomplish the Text-to-SQL task in an economical manner.
Our framework incorporates a semantic-enhanced schema to enrich database information and refine SQL queries through adaptive bias elimination and dynamic execution adjustment.
Specifically, we augment the schema with question-related column values to enhance its semantic understanding. We then fine-tune and employ a bias eliminator to mitigate the inherent biases of the LLM during SQL generation. With dynamic execution adjustment, it ensures the executability of SQL queries by interleaving reflection and correction within the execution process.

We evaluate the effectiveness of our approach by conducting evaluations on multiple datasets, including the Spider dataset and its variant dataset, as well as the BIRD dataset. Our framework demonstrates remarkable performance on the Spider dataset, achieving an accuracy of 83.6\% on the development set. This surpasses the state-of-the-art (SOTA) GPT-3.5-based method and is competitive to the performance of GPT-4. In addition, when testing on the Spider variant dataset, specifically Spider-Realistic development set, our framework exhibits outstanding performance.
Furthermore, on the BIRD dataset, our framework achieves an execution accuracy of 56.13\% on the development set, outperforming all GPT-3.5-based methods with only 9\%-58\% generation cost. This result is on par with the performance of other GPT-4-based methods with only 0.9\%-5.3\% generation cost, highlighting the competitive capabilities of our approach.
The main contributions of our work are as follows:

\begin{itemize}
    \item We propose an economically efficient and highly effective framework SEA-SQL for leveraging LLMs to accomplish Text-to-SQL Task with zero-shot prompts.
    \item  We propose adaptive bias elimination and dynamic execution adjustment to mitigate potential errors in SQL queries and improve the performance of the Text-to-SQL task.
    \item SEA-SQL significantly outperforms previous SOTA GPT-3.5 based Text-to-SQL method, and obtains an excellent performance competitive to GPT-4 with lower cost.
\end{itemize}

\begin{figure*}[ht]
	\centering 
	\includegraphics[width=\textwidth]{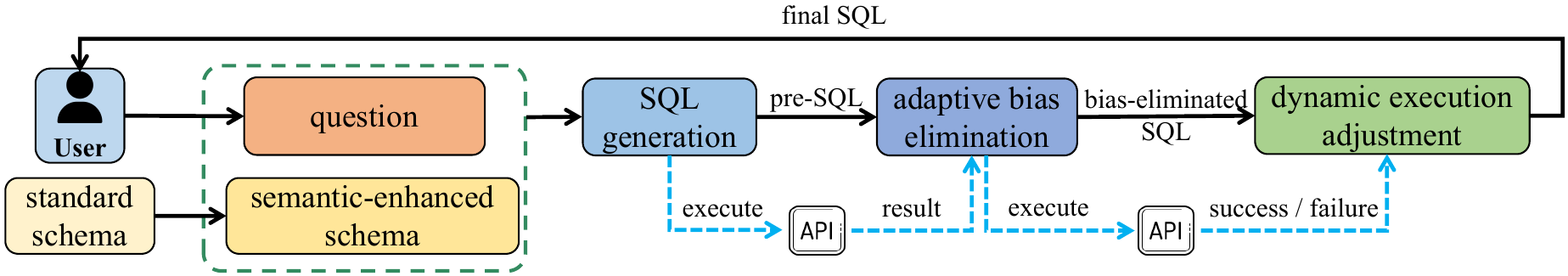}	
	\caption{An overview of the SEA-SQL framework: The SQL query is generated using the semantic-enhanced schema. Then, through adaptive bias elimination, the inherent biases caused by the LLM are eliminated. Finally, with dynamic execution adjustment, the SQL query is ensured to execute successfully to return the expected results to the user.}
    \label{image1}
    \vspace{-0.2cm}
\end{figure*}

\section{Related Work}

With the rise of LLMs, individuals are progressively exploring the applications of LLMs, achieving successful outcomes particularly in the Text-to-SQL domain. For example, ChatGPT's Evaluation \cite{DBLP:journals/corr/abs-2303-13547} conducts a comprehensive evaluation about zero-shot performance of GPT-3.5's ability in Text-to-SQL task. C3 \cite{DBLP:journals/corr/abs-2307-07306} proposes a zero-shot method utilizing GPT-3.5 to generate SQL, improving the generated SQL by providing clear layouts, calibrating bias prompts, and enhancing it through self-consistency \cite{DBLP:conf/iclr/0002WSLCNCZ23}. DAIL-SQL \cite{DBLP:journals/pvldb/GaoWLSQDZ24} investigates various prompt representations and few-shot example selection methods to determine the most effective approach for efficient SQL generation. DIN-SQL \cite{DBLP:conf/nips/PourrezaR23} divides SQL generation into three stages: schema linking, SQL difficulty decomposition, and post-processing to refine SQL. MAC-SQL 
\cite{DBLP:journals/corr/abs-2312-11242} transforms the task of Text-to-SQL into a multi-agent collaboration task, constructs a selector for schema linking, a decomposer to break down questions and generate SQL, and a refiner to ensure executable SQL.

However, there are some issues with the methods mentioned earlier. Firstly, their ability to eliminate model bias is restricted to the model itself. This is achieved either by constraining model generation via prompts \cite{DBLP:journals/corr/abs-2307-07306}, or by creating rule-based prompts to correct SQL \cite{DBLP:conf/nips/PourrezaR23, DBLP:journals/corr/abs-2312-11242}. Since the model may not always adhere to prompts or recognize its own deficiencies, this strategy has limitations in eliminating model bias. Secondly, while MAC-SQL proposes ensuring SQL executability through execution, it identifies only five types of errors that the model may encounter in the prompt based on specific error cases. Consequently, the model's effectiveness in addressing unknown errors is restricted, thereby limiting the model's overall capability. Thirdly, most methods utilize GPT-4 for SQL generation, which comes with expensive costs. Although C3 \cite{DBLP:journals/corr/abs-2307-07306} exclusively utilizes GPT-3.5, it suffers from the drawback of inferior performance.

Additionly, DAIL-SQL \cite{DBLP:journals/pvldb/GaoWLSQDZ24} presents a method to fine-tune open-source LLMs for the Text-to-SQL task. However, their performance is notably inferior, displaying a significant gap compared to other models in the GPT series.

\section{Task Definition}

Given a question $Q_{qes}$ and a database schema $S$ consisting of:
\begin{enumerate}
    \item A set of tables $T = { t_1, t_2, \dots, t_N }$, where $N$ is the total number of tables. Each $t_i$ represents the $i$-th table.
    \item A set of columns $C = { c^1_1, \dots, c^1_{n_1}, \dots, c^N_1, \dots,}$ ${c^N_{n_N} }$, where $n_i$ denotes the number of columns in table $t_i$, and $c^i_j$ represents the $j$-th column of table $t_i$.
    \item A set of values $V = {v^1_1, \dots, v^1_m, \dots, v^{M}_1, \dots, v^{M}_m }$, where $M = \sum_{i=1}^{N} n_i$ is the total number of columns. Each column has $m$ sample values, and $v^j_k$ denotes the $k$-th value of column $c^j$.
    \item A set of foreign keys $R = \left \{ \left ( t_i.c_k = t_j.c_l \right ) \right \} $, where $ t_i, t_j \in T, c_k, c_l \in C $.
\end{enumerate}

The Text-to-SQL task is to generate an SQL $A$ based on the question $Q_{qes}$ and the schema $S$. This generated SQL can be executed on the database to get the answer of the question $Q_{qes}$.

\section{Methodology}

\begin{figure*}[ht]
	\centering 
	\includegraphics[width=\textwidth]{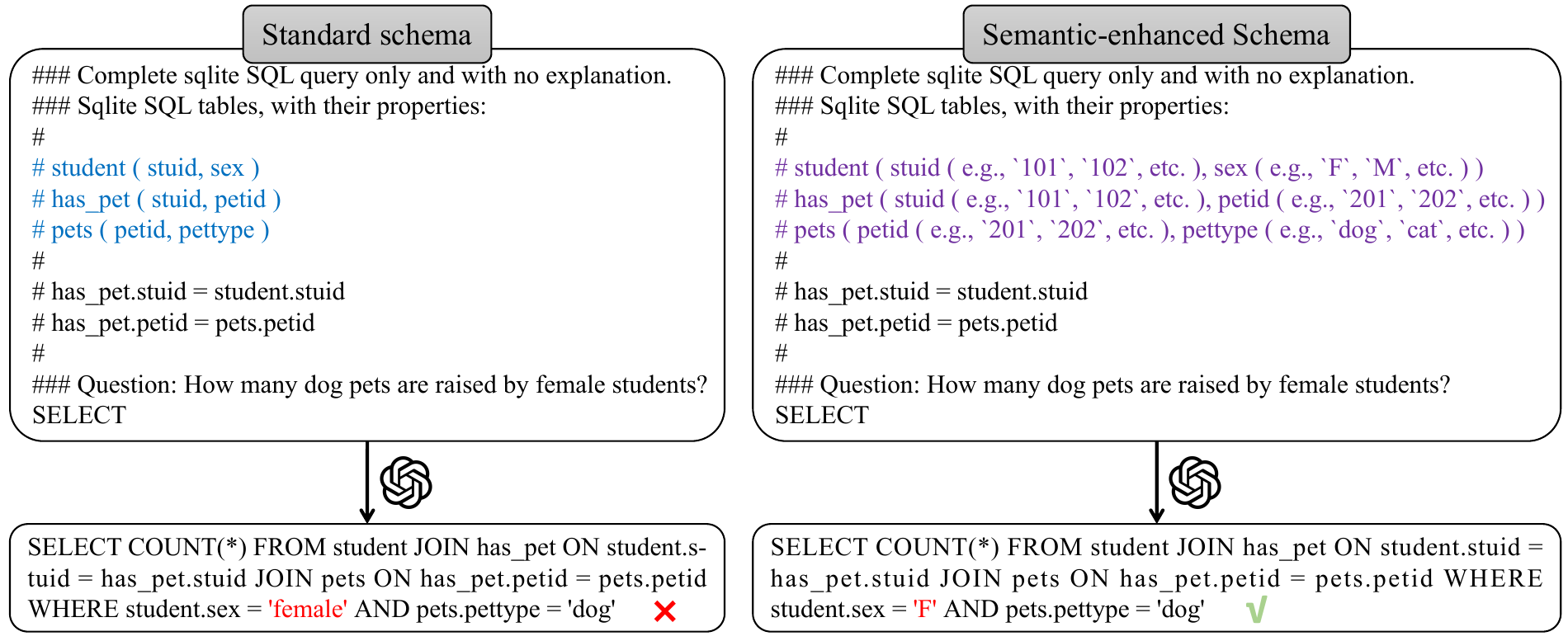}	
	\caption{Comparison of different schema formats. The left shows a standard schema, while the right showcases a semantic-enhanced schema.
 }
    \label{image2}
    \vspace{-0.2cm}
\end{figure*}

To optimize the Text-to-SQL task by LLMs and ensure the generation of executable SQL that effectively produces the correct results, we propose SEA-SQL. As illustrated in Figure \ref{image1}, our framework incorporates a semantic-enhanced schema to enrich the database information and refine SQL queries through adaptive bias elimination and dynamic execution adjustment with zero-shot prompts.

Specifically, the semantic-enhanced schema includes column values related to the question, which facilitates the LLM's identification and utilization of the appropriate tables and columns during SQL query generation. Subsequently, an adaptive bias eliminator, which is a fine-tuned small LLM (e.g., Mistral-7B), is used to eliminate the inherent biases of the LLM within the SQL queries, thereby enhancing the quality of the generated SQL queries. Finally, with dynamic execution adjustment, an iterative process of SQL execution and LLM reflection-correction is employed until the SQL query is executed successfully.

\subsection{Semantic-enhanced Schema}

The study \cite{DBLP:conf/emnlp/LinSX20} points out the limitations of solely relying on table and column names, along with their foreign key relations, to grasp the semantics of a database schema with respect to a specific question. It is demonstrated that such an approach potentially overlooks crucial contextual information contained within the actual data values of the database. Therefore, it becomes necessary to integrate the actual values present in the database into the SQL generation process to mitigate this limitation. While previous research \cite{DBLP:conf/emnlp/LinSX20} suggests augmenting the schema with values extracted from the columns mentioned in the question, this approach encounters challenges, particularly when the question does not explicitly specify the necessary values. For instance, consider a question like "the average salary of America", where the "country" column only contains the value "USA". In such cases, without supplemental information, it is difficult to accurately deduce the intended value format of the question.

To enable the LLM to understand the database comprehensively and utilize table names, column names, and value information correctly, we propose a simple yet effective method \cite{DBLP:journals/dase/WanH24}.
Although the "bridging" method \cite{DBLP:conf/emnlp/LinSX20} yields excellent performance, its application is limited by slow processing speeds when handling large datasets. Additionally, it extracts only the values explicitly mentioned in the question, which can result in insufficient information capture. To address these limitations, we propose a new retrieval paradigm that combines BM25 \cite{DBLP:journals/pacmmod/LiZLFZZWP0024} with the "bridging" method \cite{DBLP:journals/fcsc/WangZTH23}. In this integrated approach, "bridging" functions as a second-stage retrieval mechanism, enhancing the overall efficiency and depth of information retrieval.

Initially, we construct a BM25 index for each column of the database, leveraging the inherent structure and relationships within the data. Using this index, we retrieve a subset of the top-k sample values (e.g., k=1000) from each column in the database.
We then apply the "bridging" method \cite{DBLP:conf/emnlp/LinSX20} for string matching. If any of the top-k values perfectly match specific words in a given question, these values are moved to the new top-1 position. Finally, we select the top-m values (e.g., m=2) from these rearranged lists as sampled values.
We apply these sampled values to enhance the semantic representation of the database schema. As shown in Figure \ref{image2}, compared to a standard schema, semantic-enhanced schema retrieves the top-m values of columns relevant to the question, assisting the LLM in better understanding the database content. This ensures that even in cases of ambiguous column names or particularly unique format column values, the LLM can still comprehend the database information accurately, enabling correct usage of table names, column names, and values.

The LLM then generates an SQL query based on both the original question $Q_{qes}$ and the database schema enhanced with semantic information $S$, following the zero-shot prompt format proposed by \cite{DBLP:journals/corr/abs-2303-13547}, as detailed in Appendix \ref{sec:appendix}:
\begin{equation}
A_{pre} = \text{LLM}_\text{generate} \left ( Q_{qes}, S \right )
\end{equation}
where $A_{pre}$ represents the preliminary SQL query (pre-SQL) generated by the LLM with semantic-enhanced schema. This pre-SQL serves as the foundation for the subsequent steps in SEA-SQL.

\subsection{Adaptive Bias Elimination}

Due to inherent biases, LLMs may generate SQL queries that contain specific model-related errors. These errors often occur because the models tend to favor certain patterns influenced by biases present in the training data, e.g., GPT-3.5 prefer to use "LEFT JOIN" rather than "JOIN" \cite{DBLP:journals/corr/abs-2303-13547, DBLP:journals/corr/abs-2307-07306}. While approaches like DIN-SQL \cite{DBLP:conf/nips/PourrezaR23} suggest using specific prompts to address biases, they are based on few-shot examples and manually add rules in prompts from the error case analysis, which may not cover all potential error cases and may not correct the errors effectively. To comprehensively eliminate the errors caused by the LLM inherent biases, we propose an adaptive bias eliminator, which adaptively discovers and corrects the inherent biases caused by the SQL generation. 

$\bullet$ \textbf{Objective}. Previous study \cite{DBLP:journals/pvldb/GaoWLSQDZ24} indicates that a fine-tuned small LLM (e.g., Mistral-7B) does not yield comparable performance to GPT-3.5 or GPT-4 for directly generating SQL queries. The challenges encountered by the small LLM is attributed to the complexities involved in understanding SQL syntax and database information. In this paper, we propose a simplified task for the small LLM, suggesting it act as a bias eliminator of potential biases that might be introduced by the large SQL generation model (e.g., GPT-3.5). Therefore the small LLM does not need to fully comprehend all aspects of SQL syntax or complex database information. The primary role transition to identifying and eliminating inherent biases in the SQL generation model, ensures the production of high-quality and unbiased SQL queries.

Specifically, the bias eliminator receives two types of information: (1) SQL queries generated by the SQL generation model (pre-SQL), and
(2) the execution results of queries include the execution information (success or failure) and their corresponding results.
The former provides the coarse-grained information for the bias eliminator, while the latter offers a fine-grained information of the generated SQL queries, helping in the detection of potential biases in the generated SQL queries. The bias eliminator then outputs a revised SQL query, which removes any identified biases:
\begin{equation}
A_{elim} = {M}_{elim} \left ( Q_{qes}, S, A_{pre}, r \right )
\end{equation}
where $A_{elim}$ denotes the bias-eliminated SQL, $M_{elim}$ represents the adaptive bias eliminator, and $r$ stands for the corresponding execution result of $A_{pre}$. If an error occurs during execution and no results are obtained, the content of $r$ will be the error message.

\begin{figure}[t]
    \centering
    \includegraphics[width=\linewidth]{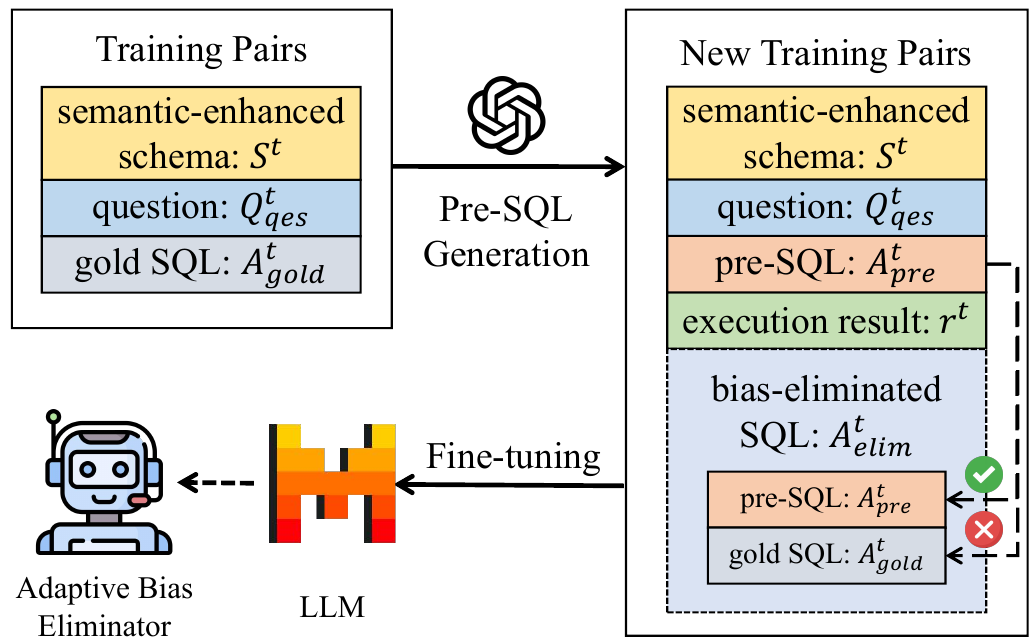}	
    \caption{The fine-tuning process of Adaptive Bias Elimination. A checkmark (\checkmark) indicates that the execution result of \( A^{t}_{\text{pre}} \) is identical to the execution result of \( A^{t}_{\text{gold}} \), while a crossmark (\crossmark) indicates they differ.}
    \label{image3}
    \vspace{-0.2cm}
\end{figure}

$\bullet$ \textbf{Fine-tuning}. As shown in Figure \ref{image3}, to support the adaptive bias eliminator in identifying and mitigating biases introduced by the SQL generation LLM effectively, it must be exposed to a substantial number of bias cases produced by the SQL generation LLM. To obtain such training data, we conduct an extension to the standard training pairs: $(Q^{t}_{qes}, S^{t}, A^{t}_{gold})$, where $t$ indicates the training data, $Q^{t}_{qes}$ represents the question in the train set and $A^{t}_{gold}$ denotes the correct SQL query corresponding to the question $Q^{t}_{qes}$ and $S^{t}$ represents the semantic-enhanced schema in the train set. Concretely, we use the SQL generation LLM to generate the pre-SQL $A^{t}_{pre}$ for each question, and obtain the execution result $r^{t}$ for each $A^{t}_{pre}$. The supplementary data $r^{t}$ provides the bias eliminator with a comprehensive understanding of where the SQL generation model is prone to making errors because of the inherent model biases. 

We expect the adaptive bias eliminator to be capable of modifying SQL queries that contain model bias, while preserving the original format of queries that are free from such bias. If a pre-SQL query yields execution results identical to those of a gold SQL query, then the bias eliminator should leave the pre-SQL query unchanged, denoted as $A^{t}_{elim} = A^{t}_{pre}$. Conversely, if the pre-SQL query execution results deviate from those of the gold SQL query, the bias eliminator should adjust the query to match the gold, reflected by $A^{t}_{elim} = A^{t}_{gold}$. This ensures that modifications are only made when necessary and that the integrity of bias-free queries is maintained.
Finally, the new training pairs are expressed as: $(Q^{t}_{qes}, S^{t}, A^{t}_{pre}, r^{t}, A^{t}_{elim})$.

The training objective for the small LLM bias eliminator, denoted as $M_{elim}$ is to maximize the likelihood of output $A^{t}_{elim}$ across all $T$ training pairs:
\begin{equation}
\mathcal{L}= - \sum_{}^{T} \log_{}{p_{M_{elim}} } (A^{t}_{elim}|Q^{t}_{qes}, S^{t}, A^{t}_{pre}, r^{t})
\end{equation}

\subsection{Dynamic Execution Adjustment}

\begin{figure}[!ht]
    \centering
    \includegraphics[width=\linewidth]{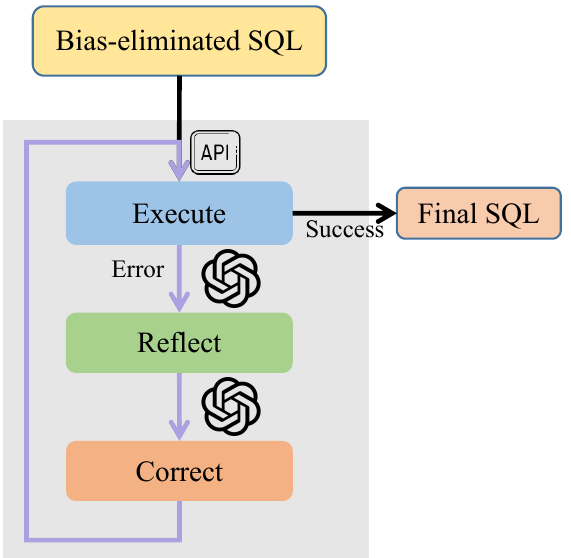}
    \caption{The process of Dynamic Execution Adjustment.}
    \label{fig:2}
    \vspace{-0.2cm}
\end{figure}

\begin{figure}[!ht]
    \centering
    \includegraphics[width=\linewidth]{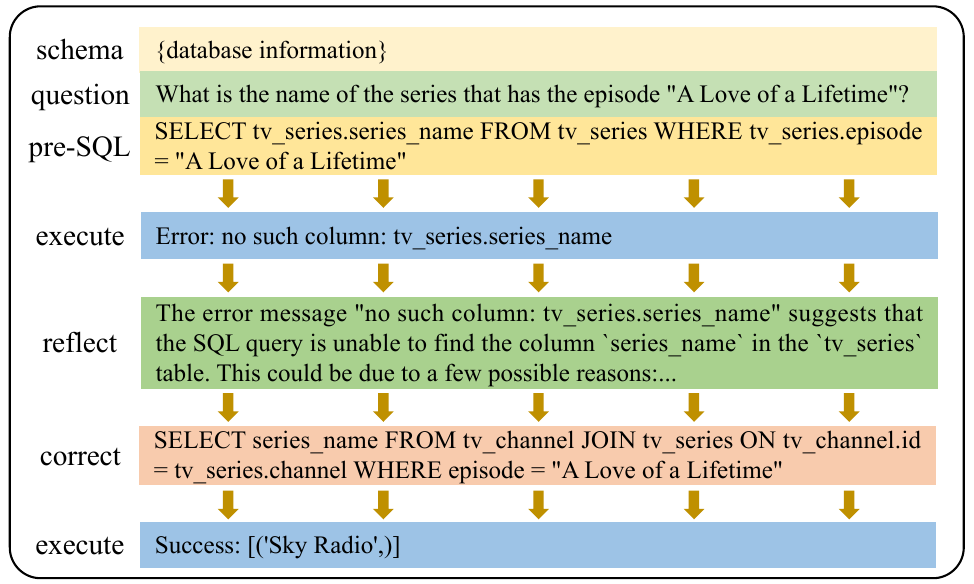}
    \caption{An example of Dynamic Execution Adjustment.}
    \label{fig:3}
    \vspace{-0.2cm}
\end{figure}

Although SQL queries may be free of biases, they can still fail to execute due to issues such as column mismatches or runtime errors. These issues are not directly related to bias, but they determine whether the SQL can ultimately execute successfully, which is a crucial aspect of the Text-to-SQL task.
However, a significant portion of prior research concentrates on the refinement of SQL exclusively through the LLM, ignoring whether the execution is successful or not \cite{DBLP:conf/nips/PourrezaR23}. As depicted in Figure \ref{fig:2}, we propose dynamic execution adjustment, which corrects unexecutable SQL queries by leveraging the LLM to both identify the error reason in the SQL query and regenerate it for execution. Inspired by Chain-of-Thought (CoT) \cite{DBLP:conf/nips/Wei0SBIXCLZ22} and REACT \cite{DBLP:conf/iclr/YaoZYDSN023}, we divide the adjustment process into three steps: execute, reflect, and correct. Through external execution of SQL API calls and continuous internal reflection and correction within the LLM, we ensure that the SQL query eventually can be successfully executed.

\textbf{Execute}. In this initial step, the SQL query is executed through external API calls to the database. This execution provides real-world feedback on the effectiveness of the SQL query. If the execution is successful, no further action is needed. However, if the execution fails, it indicates that there's an issue with the SQL query that requires correction.

\textbf{Reflect}. Once encountering a failed execution, the LLM reflects on the error encountered. This involves analyzing the error message from a failed execution and identifying the specific reason of the SQL query that caused the failure. This reflective process facilitates the LLM in understanding how the SQL query needs to be adjusted or corrected.

\textbf{Correct}. Based on the insights gained from reflection, the LLM regenerates the SQL query, thereby addressing the identified errors. This correction process involves making appropriate adjustments to the SQL syntax, structure, or logic to correct the underlying problem. The goal is to produce a revised SQL query that resolves the issues identified during reflection.

Detailed zero-shot prompts of these steps can be found in Appendix \ref{sec:appendix}. Figure \ref{fig:3} illustrates an example of dynamic execution adjustment process, the LLM dynamically corrects errors in SQL queries by utilizing error feedback and its reflection information. When an error is detected by execution, the LLM conducts a thorough reflection of its error reason and try to correct it, leading to the generation of a refined SQL query. This refined query is then undergoes another execution attempt. If additional errors emerge during this iterative process, the LLM methodically reflects upon and corrects the SQL iteratively until a version capable of successful execution is obtained.

By following the three-step approach of execute, reflect, and correct, it iteratively improves the quality and effectiveness of the generated SQL queries. Continuous feedback from the execution results and dynamic adjustments guided by the LLM ensure that the SQL queries evolve to be more accurate and executable over time. Note that, throughout the iterative process, the LLM memories previous error information, reflections, and correction results to avoid it regenerating previously erroneous SQL queries during dynamic execution adjustments.

\section{Experiment}

\subsection{Experiment Setup}

\noindent \textbf{Datasets}. The Spider dataset, as described in \cite{DBLP:conf/emnlp/YuZYYWLMLYRZR18}, is a comprehensive semantic parsing and Text-to-SQL dataset of significant scale and complexity. This dataset comprises 10,181 questions and 5,693 unique complex SQL queries. It covers 200 datasets featuring multiple tables and spans 138 diverse domains.
It comprises 8,659 instances for the training set and 1,034 instances for the development set.

The BIRD dataset, introduced in \cite{DBLP:conf/nips/LiHQYLLWQGHZ0LC23}, serves as a substantial benchmark for large-scale database tasks related to Text-to-SQL. This dataset includes 12,751 pairs of Text-to-SQL data and encompasses 95 databases, with a collective size of 33.4 GB. Covering 37 professional domains, BIRD stands out as the first Text-to-SQL benchmark to emphasize efficiency, encouraging the development of more efficient query methods tailored for massive and noisy database contents.
It comprises 9,428 instances for the training set, 1,534 instances for the development set, and 1,789 instances for the concealed test set.

To ensure the robustness of our model, we conduct a thorough evaluation using another distinct dataset: Spider-Realistic \cite{DBLP:conf/naacl/DengAMPSR21} consisting of 508 samples. This dataset is curated from the Spider dataset, wherein questions are strategically modified to emulate real-world application scenarios. Specifically, Spider-Realistic omits explicitly mentioned column names in questions.
It comprises 508 instances for the development set.

\noindent \textbf{Evaluation Metrics}.
To evaluate the performance of the Text-to-SQL task, following \cite{DBLP:conf/emnlp/ZhongYK20} and \cite{DBLP:conf/nips/LiHQYLLWQGHZ0LC23}, we adopt two metrics: Execution Accuracy (EX) and Valid Efficiency Score (VES).
\begin{itemize}
    \item The Execution Accuracy (EX) is the proportion of examples in the evaluation dataset where the execution results of the predicted SQL queries are the same as those of the gold SQL queries. For an execution result \(Y_n\) of a gold SQL denoted by \(V_n\), and an execution result of a predicted SQL \( \hat{Y}_n \) denoted by \(\hat{V}_n\), EX is computed using the formula:
    \[
    \text{EX} = \frac{\sum_{n=1}^{N} \Phi(V_n, \hat{V}_n)}{N}
    \]
    where $N$ represents the total number of SQL queries, and the indicator function \(\Phi(\cdot)\) is defined as:
    \[
    \Phi(V, \hat{V}) = 
    \begin{cases} 
    1, & \text{if } V = \hat{V} \\
    0, & \text{if } V \neq \hat{V}
    \end{cases}
    \]

    \item The Valid Efficiency Score (VES) could assess the efficiency of valid SQL queries generated by models. A "valid SQL" is defined as a predicted SQL query whose execution result matches its corresponding gold SQL result. Queries that fail to generate correct execution results are deemed invalid as they are ineffective for user needs, irrespective of efficiency. Therefore, the VES metric takes into account both the effectiveness and the efficiency of execution, providing a comprehensive measure of model performance. The VES can be calculated as follows:
    \[
    \text{VES} = \frac{\sum_{n=1}^{N} \Phi(V_n, \hat{V}_n) \cdot R(Y_n, \hat{Y}_n)}{N}
    \]
    where \( R(\cdot) \) evaluates the relative execution efficiency of the predicted SQL in relation to the gold SQL:
    \[
    R(Y_n, \hat{Y}_n) = \sqrt{\frac{E(Y_n)}{E(\hat{Y}_n)}}
    \]
    here \( E(\cdot) \) is a function that assesses the execution efficiency of each SQL in a specific setting \cite{DBLP:conf/nips/LiHQYLLWQGHZ0LC23}.
\end{itemize}

\begin{table*}[ht]
    \centering 
    \resizebox{\linewidth}{!}{\begin{tabular}{cccccc}
    \hline
    Method & Model & Zero-Shot & Few-Shot & Fine-Tune & dev  \\ \hline
    GPT-4 \cite{DBLP:conf/nips/PourrezaR23} & GPT-4 & \checkmark & & & 72.9 \\
    DIN-SQL \cite{DBLP:conf/nips/PourrezaR23} & GPT-4 & & \checkmark & & 82.8 \\
    DAIL-SQL \cite{DBLP:journals/pvldb/GaoWLSQDZ24} & GPT-4 & & \checkmark & & 83.5 \\
    MAC-SQL \cite{DBLP:journals/corr/abs-2312-11242} & GPT-4 & & \checkmark & & \textbf{86.8} \\ 
    \hdashline
    C3 \cite{DBLP:journals/corr/abs-2307-07306} & GPT-3.5 & \checkmark & & & 81.8\\
    DIN-SQL \cite{DBLP:conf/nips/PourrezaR23} & GPT-3.5 & & \checkmark & & 74.9 \\
    DAIL-SQL \cite{DBLP:journals/pvldb/GaoWLSQDZ24} & GPT-3.5 & & \checkmark & & 78.2 \\
    MAC-SQL \cite{DBLP:journals/corr/abs-2312-11242} & GPT-3.5 & & \checkmark & & 80.6 \\ \hdashline
    T5-3B \cite{DBLP:conf/emnlp/ScholakSB21} & T5-3B & & & \checkmark & 74.4 \\
    T5-3B + PICARD \cite{DBLP:conf/emnlp/ScholakSB21} & T5-3B & & & \checkmark & 79.9 \\
    Supervised Fine-Tuning \cite{DBLP:journals/pvldb/GaoWLSQDZ24}  & Llama2-7b & & & \checkmark & 66.7 \\
    Supervised Fine-Tuning \cite{DBLP:journals/pvldb/GaoWLSQDZ24}  & Llama2-13b & & & \checkmark & 67.0 \\ \hline
    SEA-SQL(ours)  & GPT-3.5 + Mistral-7B & \checkmark & & \checkmark & \underline{83.6} \\ \hline
    \end{tabular}}
    \caption{SEA-SQL and baselines on Spider development set (EX result), where the best outcomes are highlighted in bold, while the second highest results are underlined.}
    \label{table1}
    \vspace{-0.4cm}
\end{table*}

\noindent \textbf{LLMs}.
We employ two distinct LLMs for our task: GPT-3.5, a black-box model accessible solely through the API, and Mistral-7B, an open-source model that allows us not only to utilize it via the API but also affords the flexibility to freely fine-tune its parameters.
We employ GPT-3.5~\footnote{We use gpt-3.5-turbo-16k-0613 to conduct our experiment.} for SQL query generation and dynamic execution adjustment. To mitigate biases in GPT-3.5, we utilize a fine-tuned Mistral-7B model as our adaptive bias eliminator.

\noindent \textbf{Implementation Details}.
The Mistral-7B model is fine-tuned using LoRA \cite{DBLP:journals/corr/abs-2309-14717}, deepspeed \cite{DBLP:conf/sc/AminabadiRALLZRSZRH22} and flash attention \cite{DBLP:conf/nips/DaoFERR22}, employing a batch size of 64 and a learning rate of 2e-4. The training is conducted on the Spider and BIRD training datasets for one epoch using 8 × 32G V100 GPUs in 1.5 hours. Additionally, we incorporate a linear warm-up with 10\% of the training steps and employ linear decay to adjust the learning rate during training.
For semantic-enhanced schema, we retrieve the top 1000 sample values and retain the top 2 after "bridging".
To prevent the system from entering an infinite loop, we have capped the maximum iterations for dynamic execution adjustment at 10.

\noindent \textbf{Baselines.}
We conducted experiments on both the BIRD and Spider datasets, comparing our method with the following baselines:
\begin{itemize}
    \item \textbf{GPT-4 \cite{DBLP:conf/nips/PourrezaR23}} utilizes a straightforward zero-shot Text-to-SQL prompt for SQL generation.
    \item \textbf{C3 \cite{DBLP:journals/corr/abs-2307-07306}} employs clear prompting, calibration with hints, and ensures consistent output through self-consistency to generate SQL queries.
    \item \textbf{DIN-SQL \cite{DBLP:conf/nips/PourrezaR23}} decomposes the generation task into smaller sub-problems. The solutions to these sub-problems are then inputted into LLMs to produce the final SQL output.
    \item \textbf{DAIL-SQL \cite{DBLP:journals/pvldb/GaoWLSQDZ24}} leverages few-shot prompting by retrieving examples from the training set to facilitate the generation of SQL queries.
    \item \textbf{MAC-SQL \cite{DBLP:journals/corr/abs-2312-11242}} completes the text-to-SQL task through a multi-agent collaboration approach, involving schema selection, question decomposition, and SQL refinement.
    \item \textbf{T5-3B \cite{DBLP:conf/emnlp/ScholakSB21}} fine-tunes the T5-3B model for text-to-SQL tasks.
    \item \textbf{T5-3B + PICARD \cite{DBLP:conf/emnlp/ScholakSB21}} fine-tunes the T5-3B model for text-to-SQL tasks, with PICARD applying syntax rules to ensure the generated SQL is parsed incrementally, verifying each step against predefined syntax and structural constraints.
    \item \textbf{Supervised Fine-Tuning \cite{DBLP:journals/pvldb/GaoWLSQDZ24}} involves directly fine-tuning LLMs specifically for the text-to-SQL task.
\end{itemize}

\subsection{Overall Performance}

\begin{table*}[ht]
    \centering
    \resizebox{\linewidth}{!}{\begin{tabular}{ccccccccc}
    \hline
    \multirow{2}{*}{Method} & \multirow{2}{*}{Model} & \multirow{2}{*}{Zero-Shot} & \multirow{2}{*}{Few-Shot} & \multirow{2}{*}{Fine-Tune} & \multicolumn{2}{c}{dev} & \multicolumn{2}{c}{test} \\ \cline{6-9} 
     & & & & & EX & VES & EX & VES \\ \hline
    GPT-4 & GPT-4 & & \checkmark & & 46.35 & 49.77 & 54.89 & 60.77 \\
    DIN-SQL & GPT-4 & & \checkmark & & 50.72 & 58.79 & 55.90 & 59.44 \\
    DAIL-SQL & GPT-4 & & \checkmark & & 54.76 & 56.08 & \underline{57.41} & \underline{61.95} \\
    MAC-SQL & GPT-4 & & \checkmark & & \textbf{57.56} & 58.76 & \textbf{59.59} & \textbf{67.68} \\ \hdashline
    DIN-SQL & GPT-3.5 & & \checkmark & & 44.92 & 46.73 & - & - \\
    DAIL-SQL & GPT-3.5 & & \checkmark & & 44.65 & 46.99 & - & - \\
    MAC-SQL & GPT-3.5 & & \checkmark & & 50.56 & \textbf{61.25} & - & - \\ \hline
    SEA-SQL(ours)  & GPT-3.5 + Mistral-7B & \checkmark & & \checkmark & \underline{56.13} & \underline{60.83} & 53.66 & 57.00 \\\hline
    \end{tabular}}
    \caption{SEA-SQL and baselines on BIRD.}
    \label{table2}
    \vspace{-0.4cm}
\end{table*}

\begin{table*}[h]
    \centering
    \resizebox{\linewidth}{!}{\begin{tabular}{lccccccccc}
    \toprule
    \multirow{3}{*}{Method} & \multicolumn{8}{c}{BIRD} & \multirow{2}{*}{Spider} \\
    \cmidrule(lr){2-9}
    & \multicolumn{4}{c}{EX} & \multicolumn{4}{c}{VES} \\
    \cmidrule(lr){2-5} \cmidrule(lr){6-9} \cmidrule(lr){10-10}
    & Simple & Mod. & Chall. & All & Simple & Mod. & Chall. & All & EX \\ 
    \midrule
    SEA-SQL & \textbf{64.11} & \textbf{47.74} & \textbf{31.94} & \textbf{56.13} & \textbf{69.63} & \textbf{51.71} & \textbf{33.72} & \textbf{60.83} & \textbf{83.6} \\  
    \hdashline
    \quad w/o ABE & 62.49 & 45.81 & 30.56 & 54.43 & 66.45 & 49.98 & 31.32 & 58.16 & 81.8 \\ 
    \quad w/o DEA & 62.27 & 43.44 & 27.28 & 53.32 & 67.80 & 46.72 & 29.64 & 57.83 & \textbf{83.6} \\ 
    \quad w/o ABE \& DEA & 59.57 & 37.85 & 21.53 & 49.41 & 63.98 & 41.03 & 22.50 & 53.13 & 80.9 \\
    \bottomrule
    \end{tabular}}
    \caption{The results of ablation studies on the development set.}
    \label{ablation}
    \vspace{-0.2cm}
\end{table*}

\begin{table*}[h]
    \centering
    \begin{tabular}{lccccccccc}
    \toprule
    \multirow{3}{*}{Values Num} & \multicolumn{8}{c}{BIRD} & \multirow{2}{*}{Spider} \\
    \cmidrule(lr){2-9}
    & \multicolumn{4}{c}{EX} & \multicolumn{4}{c}{VES} \\
    \cmidrule(lr){2-5} \cmidrule(lr){6-9} \cmidrule(lr){10-10}
    & Simple & Mod. & Chall. & All & Simple & Mod. & Chall. & All & EX \\ 
    \midrule
    zero & 49.51 & 30.11 & 21.53 & 41.00 & 55.65 & 30.20 & 22.13 & 44.79 & 76.4 \\ 
    one & \underline{59.24} & 36.34 & \textbf{22.22} & \underline{48.83} & \underline{63.55} & 38.47 & \textbf{22.53} & \underline{52.10} & \underline{79.5} \\ 
    two & \textbf{59.57} & \textbf{37.85} & \underline{21.53} & \textbf{49.41} & \textbf{63.98} & \textbf{41.03} & \underline{22.50} & \textbf{53.13} & \textbf{80.9} \\
    three & 59.14 & \underline{36.99} & 20.83 & \underline{48.83} & 61.97 & \underline{40.88} & 23.29 & 51.95 & 78.5 \\
    \bottomrule
    \end{tabular}
    \caption{The results of the number of different sample values on the development set.}
    \label{values}
    \vspace{-0.4cm}
\end{table*}

\textbf{Spider Result}. Table \ref{table1} presents the performance comparison between our proposed framework and baseline approaches on the Spider dataset's development set. Notably, it's essential to mention that Spider has closed its test set for evaluation. Leveraging GPT-3.5, SEA-SQL outperforms other methods that either use GPT-3.5 or are fine-tuned, achieving a high EX score on the development set.

A distinguishing factor of our approach is the use of zero-shot prompts, as opposed to the few-shot prompts that is predominantly used in other methods. This highlights the simplicity and effectiveness of our approach.

Compared to fine-tuned approaches \cite{DBLP:journals/pvldb/GaoWLSQDZ24}, the utilization of a small LLM as a bias eliminator is more effective than directly employing a small LLM to generate SQL.
In comparison to GPT-3.5 based approaches, SEA-SQL significantly outperforms them, emphasizing the effectiveness of combining GPT-3.5 with fine-tuning as a promising direction for further exploration.
Moreover, the results remain competitive with DIN-SQL and DAIL-SQL in the scenario of GPT-4, demonstrating SOTA performance based on GPT-3.5 and maintaining competitiveness relative to GPT-4.

\textbf{BIRD Result}. Table \ref{table2} showcases the performance of our framework on the BIRD dataset. SEA-SQL demonstrates substantial performance superiority over other GPT-3.5-based methods. Particularly noteworthy is its competitiveness with DIN-SQL and DAIL-SQL on the development set. Our approach surpasses the current best GPT-3.5-based MAC-SQL EX result by a significant margin of 11.02\% on the development set, thereby establishing a new SOTA performance based on GPT-3.5.

\subsection{Ablation Study}

We conducted a series of ablations to investigate the pivotal components of our framework. Four sets of ablation experiments were performed on the BIRD development set, each targeting a specific aspect of our approach: the adaptive bias elimination (ABE), dynamic execution adjustment (DEA), and both the ABE and DEA.

Table \ref{ablation} presents the experimental results of the ablation study. Additionally, it illustrates the model's performance across varying complexities of SQL problems, categorized as Simple, Moderate (Mod.), and Challenging (Chall.). These categories are defined by human experts and consider factors such as Question Understanding, Knowledge Reasoning, Data Complexity, and SQL Complexity \cite{DBLP:conf/nips/LiHQYLLWQGHZ0LC23}.

The ablation study on the BIRD dataset underscores the importance of both the adaptive bias elimination and the dynamic execution adjustment components in handling SQL queries of varying complexity. Removing either component significantly reduced performance, and eliminating both drastically worsened the outcomes. The impact of removing these components was particularly severe for moderate and complex SQL queries, while simpler SQLs experienced a smaller decline in performance. This suggests that our proposed SEA-SQL framework is particularly beneficial for generating more complex SQL queries.
In the case of the Spider dataset, the DEA component did not influence the final results due to the dataset's relative simplicity. All generated bias-eliminated SQL queries were executable, rendering the dynamic execution adjustment superfluous in improving performance.

\begin{table*}[h]
    \centering
    \resizebox{\linewidth}{!}{
    \begin{tabular}{ccccc|ccc}
    \hline
    Method & Model & Pre (\$) & Generate / Query (\$) & All (\$) & GPU-Time / Query (s) & API-Time / Query (s) & Full-Time / Query (s) \\ \hline
    GPT-4 & GPT-4 & - & 0.0351 & 54 & - & 1.17 & 1.17 \\ 
    DIN-SQL & GPT-4 & - & 0.7867 & 1207 & - & 13.86 & 13.86 \\ 
    DAIL-SQL & GPT-4 & - & 0.1232 & 189 & - & 12.29 & 12.29 \\ 
    MAC-SQL & GPT-4 & - & 0.2151 & 330 & - & 15.84 & 15.84 \\ \hdashline
    DIN-SQL & GPT-3.5 & - & 0.0776 & 119 & - & 13.86 & 13.86 \\ 
    DAIL-SQL & GPT-3.5 & - & \underline{0.0113} & \textbf{17} & - & 12.29 & 12.29 \\ 
    MAC-SQL & GPT-3.5 & - & 0.0209 & \underline{32} & - & 15.84 & 15.84 \\  \hline
    SEA-SQL & GPT-3.5 & 44.78 & \textbf{0.0065} & 55 & 1.09 & 7.37 & 8.46 \\  \hline
    \end{tabular} }
    \caption{The cost of different methods on the BIRD development set.
    }
    \label{cost}
    \vspace{-0.2cm}
\end{table*}

Overall, the results definitely underscore the crucial role played by each component. Notably, the absence of any module consistently resulted in a noticeable decrease in overall performance. This highlights the interdependence of the modules within our framework and underscores their collective significance in attaining optimal results.

\subsection{Results on Different Sample values}

To investigate the impact of the semantic-enhanced schema on performance, we conducted comparative analyses between the standard schema and the semantic-enhanced schema, utilizing varying numbers of sample values. Given that modifying the number of samples necessitates retraining of the adaptive bias eliminator, our evaluation focused solely on the performance of pre-SQL.
Due to the excessive number of sample values, the input may become too lengthy, making it challenging for the LLM to accurately comprehend the database information. Consequently, we conducted tests using 0, 1, 2, and 3 samples, as illustrated in the Table \ref{values}. The results provided significant insights into the importance of sample values in the effectiveness of our methodology.

When the number of values is zero, essentially using standard schema, a notable 17\% decrease is observed. This reduction underscores the significance of our semantic-enhanced schema in achieving optimal results.

While using only two sample values yielded the best results exclusively for the moderate complexity level of the BIRD dataset, this approach closely approached optimal performance for both the simple and complex complexity levels. Additionally, the average performance achieved by using two sample values surpassed that obtained with either one or three sample values across both the BIRD and Spider datasets.
This improvement likely occurs because two values offer more informative content than a single value, while three values can make the input overly long, potentially degrading performance.

\subsection{Cost Analysis}

One primary concern in the development of Text-to-SQL technology is its accessibility to ordinary users \cite{DBLP:journals/fcsc/JiangNLBJW24, DBLP:journals/fcsc/JinCLLW24, DBLP:journals/fcsc/LiZCLRDWZY23}. To address this concern, we conducted an evaluation of our approach, comparing it against established baselines in terms of cost effectiveness (up to May 2024)~\footnote{We use gpt-3.5-turbo-16k-0613 to calculate the cost of GPT-3.5 and gpt-4(8k) to calculate the cost of GPT-4.}, as outlined in Table \ref{cost}, where "Pre" denotes the cost of pre-processing, "Generate" refers to the cost required to generate each SQL query, and "All" represents the total cost when generating all SQL queries.
"GPU-Time" denotes the duration each query utilizes GPU, "API-Time" refers to the API request time for each SQL query, and "Full-Time" indicates the total time required to generate a complete SQL query.
Our findings reveal a significant cost advantage with our framework over the GPT-4 based approach. Since SEQ-SQL only uses 25GB of GPU and takes 1.09 seconds to generate a single SQL, this generation overhead can be considered negligible. Although our GPT-3.5 model incurs slightly higher costs with other GPT-3.5 based methods, primarily attributed to pre-processing, the expense is significantly mitigated after training the bias eliminator. Notably, the cost of generating a single SQL query is still minimal, ranging from 9\% to 58\% compared to other GPT-3.5 based methods due to zero-shot prompts. Although our method involves executing SQL queries, it is more efficient than other approaches. The use of a zero-shot prompt significantly reduces the overall execution time by minimizing the prompt length compared to few-shot methods.

\begin{table}[t]
    \centering
    \begin{tabular}{ccc}
    \hline
    Method & Model & EX \\ \hline
    DIN-SQL & GPT-4 & \underline{78.1} \\
    DAIL-SQL & GPT-4 & 76.0 \\
    MAC-SQL & GPT-4 & \underline{78.1} \\ \hdashline
    C3 & GPT-3.5 & 75.4 \\
    DIN-SQL & GPT-3.5 &  65.0 \\
    DAIL-SQL & GPT-3.5 &  69.3 \\
    MAC-SQL & GPT-3.5 & 70.3 \\  \hline
    SEA-SQL(ours)  & GPT-3.5 + Mistral-7B & \textbf{80.1} \\ \hline
    \end{tabular}
    \caption{SEA-SQL and baselines on Spider-Realistic.}
    \label{robustness}
    \vspace{-0.3cm}
\end{table}

However, it's important to acknowledge the influence of dataset size on cost efficiency. With a relatively small number of queries (only 1534) in the development set, our framework's expenses appear higher. Nonetheless, when extrapolated to larger datasets, such as facing thousands or even tens of thousands of user queries, the overhead required by our approach remains minimal.

Unlike other methods based on GPT-4 that yield noteworthy results but at high costs, SEA-SQL is more affordable. It leverages GPT-3.5 and requires only 0.9\% to 5.3\% of the expenses associated with GPT-4 for generating a single SQL query, yet it achieves competitive results. Consequently, SEA-SQL emerges as more aligned with practical needs, offering a cost-effective solution without compromising performance.

\subsection{Robustness}

\begin{table*}[h]
    \centering
    \resizebox{\linewidth}{!}{\begin{tabular}{llcccccccc}
    \toprule
    \multirow{2}{*}{Generation Model} & \multirow{2}{*}{Bias Elimination Model} & \multicolumn{4}{c}{EX} & \multicolumn{4}{c}{VES} \\
    \cmidrule(lr){3-6} \cmidrule(lr){7-10}
    & & Simple & Mod. & Chall. & All & Simple & Mod. & Chall. & All \\ 
    \midrule
    GPT-3.5 & Mistral-7B & 64.11 & 47.74 & 31.94 & 56.13 & 69.63 & 51.71 & 33.72 & 60.83 \\ 
    GPT-4o-mini & Mistral-7B & \underline{65.30} & 44.73 & \underline{36.11} & 56.32 & \underline{68.52} & 46.88 & 36.70 & 58.98 \\
    Qwen-2.5-72b-it & Mistral-7B & 64.86 & \underline{49.46} & 35.42 & \underline{57.43} & 68.22 & \underline{58.20} & \underline{38.04} & \underline{60.53} \\
    Llama-3.1-70b-it & Mistral-7B & \textbf{66.70} & \textbf{53.76} & \textbf{47.22} & \textbf{60.95} & \textbf{70.45} & \textbf{62.30} & \textbf{51.41} & \textbf{66.19} \\
    GPT-3.5 & Gemma2-7B & 62.70 & 48.17 & 33.33 & 55.54 & 68.05 & 51.78 & 33.54 & 59.88 \\ 
    GPT-3.5 & Llama2-7B & 62.59 & 43.66 & 31.25 & 53.91 & 67.19 & 47.10 & 32.20 & 57.82 \\
    \bottomrule
    \end{tabular}}
    \caption{The results of different backbones on the BIRD development set.}
    \label{backbone}
    \vspace{-0.3cm}
\end{table*}

We further assessed the robustness of SEA-SQL by validating it on the Spider-realistic dataset. In particular, we conducted experiments on the untrained Spider-realistic dataset, the results of which are detailed in Table \ref{robustness}. Remarkably, our framework showcases superior performance even when compared to the GPT-4-based approach. This underscores the robustness and generalization capabilities of our zero-shot prompts, suggesting its efficacy across diverse datasets and scenarios, thus surpassing the limitations typically associated with few-shot prompts.

\subsection{LLM Backbones}

To further verify the effectiveness of SEA-SQL, we conducted experiments across various LLM backbones.
These experiments include tests on different generation models, such as GPT-4o-mini, Qwen-2.5-72b-it (Instruct) \cite{DBLP:journals/corr/abs-2407-10671}, and Llama-3.1-70b-it \cite{DBLP:journals/corr/abs-2407-21783}, as well as various bias elimimnation models, like Gemma2-7B \cite{DBLP:journals/corr/abs-2408-00118} and Llama2-7B \cite{DBLP:journals/corr/abs-2307-09288}.
The results, presented in Table \ref{backbone}, clearly demonstrate that SEA-SQL consistently achieves excellent performance across all LLM backbones. Notably, SEA-SQL, when based on Llama-3.1-70b-it and Qwen-2.5-72b-it, even perform better than the ones with GPT-3.5 and GPT-4o-mini. However, to ensure a fair comparison, we chose GPT-3.5 as our primary experimental backbone.

\subsection{Error Analysis}

\begin{figure}[h]
    \centering 
    \includegraphics[width=0.35\textwidth]{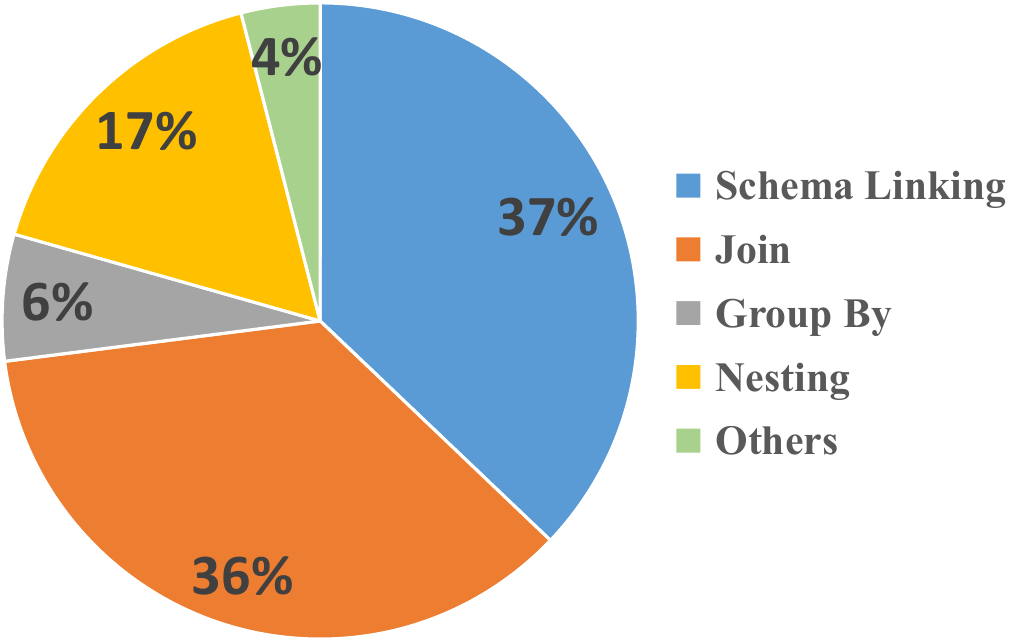}	
    \caption{Error Distributions of SEA-SQL on the BIRD development set.}
    \label{error}
    \vspace{-0.2cm}
\end{figure}

To better understand the constraints of our approach, we conducted an evaluation on the development set of the BIRD datasets. Following the methodology of DIN-SQL \cite{DBLP:conf/nips/PourrezaR23}, we classified errors into five categories: schema linking errors, JOIN errors, GROUP BY errors, nesting errors, and other errors. Schema linking errors encompass the misuse of incorrect tables, columns, or values. JOIN errors are characterized by the use of incorrect columns or tables for JOIN operations. GROUP BY errors arise when incorrect columns are used or when GROUP BY is not utilized. Nesting errors involve the application of incorrect sub-queries or improper operator usage.

As illustrated in Figure \ref{error}, schema linking errors are the most common, accounting for 37\% of errors in the BIRD dataset. Additionally, JOIN errors also represent a substantial segment, accounting for 36\% of the total errors, a percentage nearly equivalent to that of schema linking errors. Conversely, nesting and GROUP BY errors represent a minor portion of the total errors. These observations underscore two primary limitations of our approach: first, it may select incorrect tables or columns when faced with a large amount of irrelevant information; second, it struggles with handling complex SQL queries.

\section{Conclusion}

In this paper, we proposed SEA-SQL, a powerful framework that utilized GPT-3.5 to generate SQL queries and is competitive to GPT-4. We implemented this framework with a smaller LLM (Mistral-7B) to eliminate the biases inherent in the larger LLM (GPT-3.5). Additionally, we employed GPT-3.5 for dynamic execution adjustment, which reflects and corrects unexecutable SQL queries. Experimental results on two datasets demonstrated the effectiveness of our framework. By using a smaller LLM to eliminate bias, our approach yields better results compared to generating SQL queries directly.

\section{Acknowledgments}

This work is supported by the National Natural Science Foundation of China (Nos. 62272054, 62192784,62172056), Beijing Nova Program (No. 20230484319), and Xiaomi Young Talents Program.

\bibliography{anthology,custom}
\bibliographystyle{acl_natbib}

\appendix

\begin{table*}[!h]
\centering
\resizebox{\linewidth}{!}{\begin{tabular}{ccccccccc}
\hline
 & \multicolumn{4}{c}{EX} & \multicolumn{4}{c}{VES} \\
\cmidrule(lr){2-5} \cmidrule(lr){6-9} 
 & Simple & Mod. & Chall. & All & Simple & Mod. & Chall. & All \\ \hline
SEA-SQL (w/o further bias elimination) & 64.11 & 47.74 & 31.94 & \textbf{56.13} & 69.63 & 51.73 & 33.72 & \textbf{60.83} \\
SEA-SQL (w/ further bias elimination) & 63.89 & 47.31 & 31.94 & 55.87 & 69.28 & 51.63 & 32.81 & 60.50 \\ 
\hline
\end{tabular}}
\caption{The results of the further refinement on the BIRD development set.}
\label{further}
\vspace{-0.2cm}
\end{table*}

\begin{table*}[!h]
\centering
\resizebox{\linewidth}{!}{
\begin{tabular}{ccccccccccc}
\hline
 & \multicolumn{4}{c}{EX} & \multicolumn{4}{c}{VES} & \multirow{2}{*}{Generate / Query (\$)} & \multirow{2}{*}{Full-Time / Query (s)} \\
 \cmidrule(lr){2-5} \cmidrule(lr){6-9} 
 & Simple & Mod. & Chall. & All & Simple & Mod. & Chall. & All &  &   \\ \hline
SEA-SQL (0-shot) & 64.11 & 47.74 & 31.94 & 56.13 & 69.63 & 51.73 & 33.72 & 60.83 & \textbf{0.0065} & \textbf{8.46} \\
SEA-SQL (1-shot) & \underline{64.11} & \underline{47.96} & \textbf{34.03} & \underline{56.39} & \underline{70.23} & \underline{52.99} & \textbf{36.28} & \underline{61.82} & \underline{0.0100} & \underline{9.01} \\ 
SEA-SQL (3-shot) & \textbf{64.76} & \textbf{48.82} & \underline{33.33} & \textbf{56.98} & \textbf{71.03} & \textbf{54.15} & \underline{34.76} & \textbf{62.51} & 0.0130 & 13.98 \\ 
\hline
\end{tabular}
}
\caption{The results from using few-shot examples on the BIRD development set. Where "Generate" refers to the cost of generating each SQL query (\$), while "Full-T" represents the time taken to generate each SQL query (s).}
\label{few}
\vspace{-0.2cm}
\end{table*}

\begin{mdframed}[backgroundcolor=white, linecolor=white]
\textcolor{white}{\\
\\ \\ \\ \\ \\ \\ \\ \\ \\ \\ \\ \\ \\ \\ \\
\\ \\ \\ \\ \\ \\ \\ \\ \\
\\}
\end{mdframed}

\section{Further Bias Elimination}

After dynamic execution adjustment, new SQL quer-ies are generated by the LLM. To demonstrate that these queries do not introduce new biases, we conducted an experiment on the BIRD
dataset by applying further bias elimination process after the dynamic execution adjustment. The results, shown in Table \ref{further}, indicate a decline in performance after this process. This decline is primarily because the dynamic execution adjustment maintains the original style of the SQL, ensuring it executable without introducing biases or syntax errors. Further bias elimination did not improve the results and might make the SQL non-executable.

\section{Few-shot Prompts}

Previous work suggests that few-shot prompts may be more effective in leveraging LLMs for data-related tasks \cite{DBLP:conf/vldb/ZhangD0O24}. Consequently, we explored whether using few-shot prompts would yield greater improvements, as shown in Table \ref{few}. The results indicate that using a 1-shot approach achieves better outcomes compared to zero-shot, and a 3-shot approach yields even higher results. However, the computational cost significantly increases with more examples. Given these costs, we continue to use zero-shot prompts for our main experiments. However, in scenarios where accuracy is a priority, using few-shot prompts tends to be more effective.

\section{Prompts}
\label{sec:appendix}

Here we present out prompts utilized for each module, including SQL generation, adaptive bias elimination and dynamic execution adjustment.

\subsection{SQL Generation}

\begin{mdframed}[backgroundcolor=gray!20, linecolor=gray]
\#\#\# Complete sqlite SQL query only and with no explanation.\\
\#\#\# Sqlite SQL tables, with their properties:\\
\#\\
\{schema\}\\
\#\\
\#\#\# Question: \{question\}\\
SELECT
\end{mdframed}

\subsection{Adaptive Bias Elimination}

\begin{mdframed}[backgroundcolor=gray!20, linecolor=gray]
Below is an instruction that describes a task, paired with an input that provides further context. Write a response that appropriately completes the request.

\#\#\# Instruction:
This represents the SQLite SQL query that has been generated in response to the given question, along with the resulting outcome after executing the query.\\
Please judge its correctness based on the execution result and the explanation for the question.\\
If it's incorrect, output the correct sqlite SQL query; otherwise, output the original sqlite SQL query.\\
\\
\#\#\# Input:\\
\#\#\# Sqlite SQL tables, with their properties:\\
\#\\
\{schema\}\\
\#\\
\#\#\# Question: \{question\}\\
\#\#\# SQLite SQL query: \{SQL\}\\
\#\#\# Run results: \{result\}\\
\\
\#\#\# Response:
\end{mdframed}

\subsection{Dynamic Execution Adjustment}
Here "previous\_information" has the following format:
\begin{mdframed}[backgroundcolor=gray!20, linecolor=gray]
\#\#\# SQL: (error bias eliminated $SQL$)\\
\#\#\# Error message: (error message of the bias eliminated $SQL$)\\
\#\#\# Error reason: (error reason of the bias eliminated $SQL$)\\
\#\#\# New SQL: (the corrected new $SQL_1$)\\
...\\
\#\#\# Error message: (error message of $SQL_i$)\\
\#\#\# Error reason: (error reason of the $SQL_i$)\\
\#\#\# New SQL: (the corrected new $SQL_{i+1}$)\\
\#\#\# Error message: (error message of $SQL_{i+1}$)\\
\end{mdframed}

\subsubsection{Reflect}

\begin{mdframed}[backgroundcolor=gray!20, linecolor=gray]
\#\#\# Here is a sqlite SQL query that resulted from a question, but it produced an error when executed. What do you think is the possible reason for this SQL error?\\
\\
\#\#\# Sqlite SQL tables, with their properties:\\
\#\\
\{schema\}\\
\#\\
\#\#\# Question: \{question\}\\
\{previous\_information\}\\
\#\#\# Error Reason:
\end{mdframed}

\subsubsection{Correct}

\begin{mdframed}[backgroundcolor=gray!20, linecolor=gray]
\#\#\# Here is a sqlite SQL query that resulted from a question, but it produced an error when executed. Please correct it with no explanation.\\
\\
\#\#\# Sqlite SQL tables, with their properties:\\
\#\\
\{schema\}\\
\#\\
\#\#\# Question: \{question\}\\
\{previous\_information\}\\
\#\#\# Error Reason: \{reason\}\\
\#\#\# Correct SQL: SELECT
\end{mdframed}

\end{document}